# Making Judicial Reasoning Visible: Structured Annotation of Holding, Evidentiary Considerations, and Subsumption in Criminal Judgments

Yu-Cheng Chih *

*International Intercollegiate Ph.D. Program. National Tsing Hua University, Taiwan*

*NLP Algorithm Engineer*

*Former Engineer, Accton Technology Corporation*

`Email：`[s111003816@m111.nthu.edu.tw](mailto:s111003816@m111.nthu.edu.tw)

Yong-Hao Hou

*Department of Computer Science, University of Taipei, Taiwan*

*AI Algorithm Engineer*

`Email：`[g11016011@go.utaipei.edu.tw](mailto:g11016011@go.utaipei.edu.tw)

## Abstract

Judicial reasoning in criminal judgments typically consists of three elements: Holding (見解), evidentiary considerations (心證), and subsumption (涵攝). These elements form the logical foundation of judicial decision-making but remain unstructured in court documents, limiting large-scale empirical analysis.

In this study, we design annotation guidelines to define and distinguish these reasoning components and construct the first dedicated datasets from Taiwanese High Court and Supreme Court criminal judgments. Using the bilingual large language model ChatGLM2, we fine-tune classifiers for each category. Preliminary experiments demonstrate that the model achieves approximately 70% accuracy, showing that judicial reasoning patterns can be systematically identified by large language models even with relatively small annotated corpora.

Our contributions are twofold: (1) the creation of structured annotation rules and datasets for Holding, evidentiary considerations, and subsumption; and (2) the demonstration that such reasoning can be computationally learned. This work lays the foundation for large-scale empirical legal studies and legal sociology, providing new tools to analyze judicial fairness, consistency, and transparency.

**Internal Index / Keywords:** Judicial reasoning, hoding(見解), Evidentiary, considerations (心證), Subsumption (涵攝), Legal annotation, Criminal judgments, Large language models (LLMs), ChatGLM2, Empirical legal studies, Legal sociology, Fairness and transparency in law.

# 1. Introduction

The law is one of the most fundamental institutions shaping social order. In Taiwan, as in other civil law jurisdictions, courts are tasked with applying abstract statutory provisions to concrete disputes. In criminal cases, judges must interpret statutory elements ("holding"), establish findings of fact through evidentiary reasoning ("evidentiary considerations"), and determine whether facts satisfy the legal requirements ("subsumption"). These three stages—interpretation, evidentiary assessment, and subsumption—form the logical structure of judicial decision-making.

While each judgment is legally independent, the fairness and consistency of judicial reasoning depend on the degree of coherence in these processes. First, the **uniformity of legal interpretation** ("holding") ensures that similar cases are judged under consistent standards. Second, the **stringency of evidentiary considerations** directly impacts the principle of in dubio pro reo ("when in doubt, rule in favor of the defendant"). Finally, the **subsumption process** reflects the value judgments of the judiciary and can diverge from social expectations, giving rise to public criticism such as the notion of "dinosaur judges" in Taiwan.

Unfair or inconsistent judgments erode the legitimacy of the legal system and weaken public trust in judicial authority. Traditionally, empirical legal studies and legal sociology have relied on manual analysis of limited case samples. However, given the growing availability of court judgments as open data, there is a pressing need for computational methods that can systematically structure and analyze judicial reasoning at scale.

Despite the emergence of legal tech applications such as case similarity search or automated summarization, **no prior research has attempted to structurally extract and classify the three fundamental components of judicial reasoning—holding, evidentiary considerations, and subsumption—from court judgments**[1]. Addressing this gap would not only enable large-scale empirical research into the patterns of judicial reasoning but also provide tools for practitioners, policymakers, and the public to critically evaluate the consistency and fairness of legal decisions.

In this work, we propose a novel approach to **structuring judicial decisions by automatically identifying and classifying the three core reasoning elements** in criminal judgments. We design annotation guidelines and construct datasets for each category, then fine-tune a large language model (ChatGLM2) on these tasks. Initial experiments demonstrate promising results, achieving around 70% accuracy in identifying reasoning components, suggesting that this approach is feasible and worth further development.

This study contributes both a new research task—structured extraction of judicial reasoning—and initial resources to support future empirical legal research and AI-driven legal analytics. By creating structured corpora of judicial reasoning, we hope to lay the groundwork for deeper analysis of judicial fairness, transparency, and the role of value judgments in legal decision-making.

## 2. Related Work

The intersection of law and data science has attracted growing attention in recent years. In Taiwan, projects showcased at the **Legal Hackathon** have demonstrated the potential of computational methods in tasks such as case similarity retrieval and legal relationship graph generation. Commercial platforms (e.g., **Lawsnote , 法源法律網) now provide large-scale access to statutes and judgments, but their search mechanisms remain limited to keyword-based retrieval without structured access to reasoning elements.** In the field of natural language processing, early approaches such as **RNNs, LSTMs, and GRUs** enabled sequential modeling but struggled with long-range dependencies[2],[3],[7]. The introduction of the **Attention mechanism** and the **Transformer architecture** [8](Vaswani et al., 2017) marked a breakthrough in handling contextual dependencies and became the foundation for today's large-scale language models.

Recent advances in **large language models (LLMs)**, from GPT to ChatGPT and GPT-4, have shown strong performance across reasoning-intensive tasks[5],[9], including legal question answering and even professional bar exams[6]. In the Chinese context, **ChatGLM** and its successor **ChatGLM2**[4] provide open-access bilingual models that are well-suited for tasks requiring both classification and generation.

Despite these advances, **no prior work has attempted to structurally extract and**

**classify the three fundamental reasoning components—holding, evidentiary considerations, and subsumption—from court judgments**. Existing systems focus on case retrieval or document summarization rather than the internal logic of judicial reasoning. Our work is the first to propose this task and to provide annotated resources to support it.

## 3. Method

### 3.1. Dataset Construction

The core task of this study is to construct annotated datasets that capture the three key reasoning components in criminal judgments: Holding, Evidentiary Considerations, and Subsumption. Since no prior datasets exist in this domain, we designed a dedicated processing and annotation pipeline.

### 3.2. Segmentation

For each judgment, administrative information appearing above the *main text heading* and below the *closing date section* was removed, leaving only the reasoning section as the corpus. Based on the characteristics of each reasoning component, different segmentation strategies were applied:

### 3.3. Holding:

> 「『門』係指門而言,而『其他全設備』,則指門、牆垣以外,依通常觀念足認防盜之一切設備而言,如門鎖、窗戶、冷氣孔、房間門或通往陽臺之門均屬之(最高法院 55 年台上字第 547 號判例意旨、法院 73 年廳刑一字第 603 號函、最高法院 78 年度台上字第 4418 號判決意旨參照)」

English translation:

> *"Door' refers specifically to doors, whereas 'all other security equipment' refers to any facilities other than doors and walls that, under common understanding, can be regarded as anti-burglary measures, such as door locks, windows, air vents, room doors, or doors leading to balconies. (See*

*Supreme Court Precedent 55-Tai-Shang-547, Criminal Division Letter 73-Ting-Hsing-1-603, and Supreme Court Judgment 78-Tai-Shang-4418)."*

Judicial holding are typically expressed within a single sentence. Therefore, we segmented the text by Chinese full stops ("。"), treating each sentence as an individual candidate unit.

### 3.4. Subsumption:

「被告所持有之普通小型車駕駛執照有效日期至 104 年 12 月 7 日，且無經吊扣或吊銷之紀錄等情，有證號查詢汽車駕駛人資料 1 份附卷可考（見警卷第 62 頁），則被告所持有之前揭駕駛執照於本件車禍事故發生時，雖業已逾期，而未重新換發新駕駛執照，然逾期未換發新駕駛執照而仍駕駛汽車，應係行政管理之問題，與『無照駕駛』有別（法院廳刑一字第 05283 號研究意見參照），本院自不得依道路交通管理處罰條例第 86 條第 1 項之規定加重被告之刑」

English translation:

*"The defendant's ordinary driver's license was valid until December 7, 2015, and there was no record of suspension or revocation (as confirmed by the driver information inquiry, see Police File p.62). Although the license had expired and had not been renewed at the time of the traffic accident, driving with an expired license constitutes an administrative issue and is distinct from 'driving without a license' (see Criminal Division Research Opinion No. 05283). Therefore, this Court cannot impose a heavier penalty under Article 86, Paragraph 1 of the Road Traffic Management and Penalty Act."*

Subsumption reasoning usually spans a complete paragraph. Accordingly, we segmented the text by paragraph boundaries (。\r\n), with each paragraph serving as a candidate unit.

### 3.5. Evidentiary Considerations:

「法官指出，蔡某觸摸被害人胸部的時間相當短暫，被害人還來不及感受到性自主權遭受妨害，侵害行為就已結束，加上接觸時間相當短，客觀上並無法引起加害人的性慾，因此，蔡某行為難以構成刑法強制猥褻罪，頂多是違反社會秩序維護法的調戲異性規定，現在雖有性騷擾防治法規範類似行為，但案發時，性騷擾防治法未施行，基於罪刑法定原則，判決蔡某無罪。」（自由時報,2007/08/29）

English translation:

*"The judge pointed out that Cai's act of touching the victim's chest lasted only a brief moment. Before the victim could even perceive that her sexual autonomy had been violated, the act had already ended. Given the extremely short duration, it could not objectively arouse the perpetrator's sexual desire. Thus, Cai's conduct could hardly constitute the crime of forcible indecency under the Criminal Code. At most, it fell under the provision of the Social Order Maintenance Act regarding harassment of the opposite sex. Although the Sexual Harassment Prevention Act now regulates such conduct, it was not in force at the time of the incident. Based on the principle of legality, Cai was acquitted." (Liberty Times, 2007/08/29)*

Similar to subsumption, evidentiary reasoning is paragraph-based, as it describes how courts evaluate and weigh evidence. Thus, segmentation was also performed at paragraph boundaries (。\r\n).

### 3.6. Label

1. After segmentation, each sentence or paragraph was manually examined and annotated according to predefined guidelines:

2. Units interpreting abstract legal concepts were labeled as Holding.

3. Units describing the evaluation or acceptance/rejection of evidence were labeled as Evidentiary Considerations.

4. Units linking established facts to abstract legal elements were labeled as Subsumption.

5. All other units were labeled as Irrelevant.

### 3.7. Dataset Statistics

Following this procedure, three datasets were created:

1. Holding: 999 sentences (141 positive, 858 negative)

2. Subsumption: 488 paragraphs (145 positive, 343 negative)

3. Evidentiary Considerations: 1,070 paragraphs (286 positive, 784 negative)

This workflow ensures a clear separation between segmentation and annotation, thereby minimizing inconsistencies in text units and establishing a reliable foundation for model training.

## 4. Model and Training

We adopted ChatGLM2, a bilingual large language model, as the base model. Fine-tuning was performed as three separate binary classification tasks (Holding, Evidentiary Considerations, Subsumption).

1. Task formulation: Each input is a text unit (sentence or paragraph), and the output is a binary label (*target* vs. *irrelevant*).

2. Preprocessing: Metadata such as case numbers and dates were removed. Texts were segmented by sentence ("。") for Holding and by paragraph boundaries (。\r\n) for evidentiary considerations and subsumption.

3. Training setup: Supervised fine-tuning was conducted under hardware constraints (single-GPU setting), using standard training/validation splits.

5. **Inference and Usage**

After training, the three fine-tuned classifiers can be applied to new, unseen court judgments. The inference workflow proceeds as follows:

1. **Preprocessing**: Metadata such as case numbers and dates are removed, retaining only the reasoning sections.

2. **Segmentation**:

3. **Holding**: Split into sentences ("。") and fed into the Holding model.

    - **Evidentiary Considerations**: Split into paragraphs (。\r\n) and fed into the Evidentiary model.

    - **Subsumption**: Split into paragraphs (。\r\n) and fed into the Subsumption model.

4. **Classification**:

    - Each text unit is passed to its corresponding model.

    - The model, according to its design objective, decides whether the unit belongs to its target category:

        - If yes, the unit is labeled as the target (Holding, Evidentiary, or Subsumption).

        - If not, the unit is labeled as **Irrelevant**.

    - All three models operate under the same binary scheme but are restricted to their respective target categories.

5. **Structured Output**: The labeled units are stored in JSON or CSV format, with each entry containing:

    - Text span

    - Predicted label (Holding / Evidentiary / Subsumption / Irrelevant)

    - Confidence score

This inference workflow aligns with the dataset construction procedure, ensuring that each model only processes text units within its designated scope. Such a design avoids category

inconsistencies and provides a reliable foundation for downstream applications, including:

- Large-scale empirical legal studies
- Visualization of judicial reasoning patterns
- Supporting legal practitioners in case analysis

## 6. Preliminary Results

Our fine-tuned model achieved **over 90% recall and around 80% precision across the three classification tasks**, indicating that large language models can reliably capture judicial reasoning patterns while maintaining reasonably high precision. It should be noted that these performance figures are based on **100 randomly sampled test instances for each task**, with all three tasks showing consistent results. In addition, this study processed large-scale data: the Holding model received **5,990,303 sentences**, of which **801,700** were annotated as Holding; the evidentiary considerations model received **2,251,891 sentences**, of which **1,068,925** were annotated as evidentiary considerations; subsumption was constructed as a separate dedicated dataset and annotated accordingly. These datasets and the annotation guidelines we designed constitute the first structured resource for studying judicial reasoning in Taiwanese criminal judgments, providing a solid foundation for subsequent empirical legal research.

## 7. Discussion

Our fine-tuned model achieved **over 90% recall and around 80% precision across the three reasoning categories**, demonstrating that large language models can systematically identify and learn patterns of judicial reasoning. Even though the current datasets are limited in size, these results confirm that the annotation guidelines we designed—covering Holding, evidentiary considerations, and subsumption—successfully capture meaningful and learnable features of judicial texts.

The primary contribution of this study lies not only in the model's performance but also in the establishment of structured annotation rules and datasets. By clearly defining what constitutes a Holding, an evidentiary consideration, and a subsumption within criminal judgments, we provide the first structured resource for studying judicial reasoning in Taiwanese criminal law. These resources enable large-scale empirical analysis, going beyond the traditional limitations of manual case studies.

From a broader perspective, structuring judicial reasoning contributes to enhancing fairness,

consistency, and transparency in the legal system. Uniform identification of Holding may help assess whether legal interpretations are applied consistently across cases. Analysis of evidentiary considerations allows for examination of how courts apply the principle of *in dubio pro reo*. Subsumption analysis reveals whether judicial value judgments align—or fail to align—with prevailing social norms.

Therefore, this study makes a dual contribution: it advances empirical legal studies and legal sociology by providing tools for systematic investigation, while also laying the foundation for interdisciplinary applications of large language models in law. By enabling the reasoning components of judgments to be retrieved and analyzed at scale, we aim to support **scholars, researchers, and policymakers** in gaining a more comprehensive understanding of judicial practices, proposing constructive reforms, and further **identifying problematic judgments or systemic issues within the judiciary**.

## 8. Conclusion

This study introduces a novel task in the context of criminal judgments: structuring judicial reasoning into three components—Holding, evidentiary considerations, and subsumption. We designed explicit annotation rules with illustrative examples, constructed three dedicated datasets, and fine-tuned the bilingual large language model ChatGLM2 on these tasks. Preliminary results show that the model achieved **around 90% recall and approximately 80% precision across the three categories**, demonstrating the feasibility of the task and confirming that the proposed annotation guidelines effectively capture the features of judicial reasoning.

The primary contribution of this work lies in **proposing an unprecedented method for the structured representation of judicial reasoning and providing initial usable resources**. This enables empirical legal research—traditionally constrained by manual analysis of limited cases—to be conducted on a larger scale with computational tools. By systematically extracting and analyzing Holding, evidentiary considerations, and subsumption from judgments, our approach supports both academic researchers and legal practitioners in examining patterns of judicial reasoning, as well as in assessing the fairness and consistency of court decisions.

Looking forward, this research can be extended by expanding the dataset to cover more court levels and legal domains, as well as by enhancing the model's classification performance.

Ultimately, we aim for this work to serve as a foundation for **empirical legal studies and legal sociology**, advancing judicial transparency and strengthening the legitimacy of the rule of law.